\documentstyle[12pt]{article}
\begin{document}
\vskip 72pt
\centerline{\bf PHOTON SENSITIVITY OF SUPERHEATED DROP} 
\centerline{\bf AT ROOM TEMPERATURE} 
\vskip 36pt
\centerline {B.Roy, Mala Das, S.C. Roy and B.K. Chatterjee}
\vskip 12pt
\centerline{\it Department of Physics, Bose Institute, Calcutta 700 009, India}
\vskip 36pt
\begin{abstract} 
It has been reported so far that superheated drops made of R-12 at room temperature 
are sensitive to neutrons yet insensitive to photons. This property makes its 
use as one of 
the most useful neutron dosimeter. The photon sensitivity of R-12 at room 
temperature (25$^oC$) when exposed 
to 59.54 keV photons obtained from radioactive $^{241}Am$ has been noted for 
the first time in our laboratory. This discovery is important not only from 
the point of view of basic science but more important to the users of R-12 
in neutron dosimetry to take note of this in assessing the neutron dose 
correctly. This discovery also indicates a new direction of investigation with 
R-12 detector.
\end{abstract}

\vskip 36pt

Superheated drops are known to detect energetic radiations[1,2,3]. Among 
the various liquids studied, R-12 
(dichlorodifluoromethane) is found to be the most widely used material 
in neutron dosimetry in view of its capability to detect thermal to fast 
neutrons at room temperature yet insensitive to gamma rays. Apfel[4] 
observed R-12 to be insensitive to gamma rays at room temperature when 
exposed to strong $^{60}Co$ source and 
assumed its gamma ray sensitivity for photons only above 6 MeV, when its energy 
is comparable to the binding energy per nucleon thus producing neutrons, 
through photo-nuclear reactions[5]. Ing et.al.[3] also 
found R-12 to be insensitive to gamma rays when exposed to a strong 
radioactive $^{60}Co$ source.

According to the empirical relationship suggetsed  by d'Errico et. al.[6],  
superheated liquid becomes sensitive to photons at a temperature close to 
the midpoint between its boiling point and critical temperature. Therefore,
according to this relation, R-12 (b.p. -29.79$^oC$, $T_c$. 112$^oC$) should 
exhibit photon sensitivity close to about 41$^oC$. Contrary to all 
these predictions, the present work demonstrates the  
sensitivity of R12 detector to gamma rays at room temperature (25$^oC$), 
when exposed to 59.54 keV photons obtained from a radioactive $^{241}Am$. 

	In this work, R12 drops have been suspended homogeneously in a 
viscoelastic gel as has been reported elsewhere[7].
R-12 sample thus prepared, taken in a 15ml glass vial was exposed to 
 gamma rays (59.54kev) from $^{241}Am$ (0.5Ci) at an average room 
temperature of 25$^oC$ and the nucleation was observed by 
counting the number (N) of drops vaporised per minute (t). This was 
done  by detecting the pressure pulse produced due to each drop vaporisation 
with the help of a piezo-electric transducer[8], coupled to a drop counter and 
multichannel scaler (MCS). The nucleation rate (dN/dt) has been recorded 
continuously in MCS. The results presented in figure 1 shows nucleation 
rate (dN/dt) with time (t). The nucleation rate at any time is proportional 
to the flux of incoming photons and the number of drops present in the 
sample at that time. Therefore dN/dt decreases with time as 
nucleation proceeds.

	Nucleations due to background radiations and due to the presence of  
heterogeneous nucleation sites, if any, have been recorded for 65 minutes 
at the beginning of the experiment. The radioactive $^{241}Am$ was then 
placed very close to the detector (at a distance of 1.2 cm from the centre of 
the detector vial) to observe 
the nucleation prominently and the nucleation rate was noted. Nucleation 
due to background was noted again after removing the radioactive source. As 
can be seen from 
Fig. 1, there is significant increase in nucleation above background when the 
source is placed near the detector, thus confirming the gamma ray sensitivity 
of R-12 at room temperature. However, the efficiency of R-12 for detecting 
such photons at room temperature is quite small as has been observed experimentally. 
When the present source was placed at a distance greater than 2.5cm from the 
centre of the vial (containing R-12 drops) there was no nucleation. 

	In a seperate experiment, the R-12 sample was exposed to 32.5 mCi 
$^{137}Cs$ (662keV) and 0.45 mCi $^{60}Co$ (average energy of 1225 keV), 
and no nucleation was noted. This result confirms the earlier observation 
of Apfel[5] and Ing et. al.[3].

	One may ask why R-12 is sensitive to low energy photons rather 
than higher energies at room temperature. The answer lies on the nature of 
interaction of photons with matter. Gamma rays are detected by the initiation 
of vapour bubbles caused by the energy deposition of electrons produced by 
photons while passing through the 
liquid. Looking at the linear energy transfer (dE/dx) of electrons at 
these energies (from tens of keVs to close to 1 MeV), one may find that
dE/dx is larger at lower energies than higher electron energies. Therefore 
the energy deposition in the liquid is larger for lower energy photons and 
favourable for nucleation compared to higher energy photons. As the temperature of the detector sample 
increases, the degree of superheat of the liquid increases and it requires 
lesser and lesser amount of energy for nucleation of drops. This 
explanation was found to be true in a seperate experiment, where the sample 
was found to be sensitive to higher energy photons (662 keV and 1225 keV) at 
higher temperatures. Complete investigation on this subject will be reported 
elsewhere.   

The present discovery of the photon sensitivity of R-12 at room temperature 
opened up a new vista of investigation. More importantly this 
discovery demonstrates that it is wrong to assume that R-12 is insensitive 
to photons at room temperature and constitutes a valid warning to the 
users of R-12 as neutron dosimeter while assessing the actual neutron dose. 
Further investigations on the sensitivity of R-12 to lower energy photons are 
needed. 

\vskip 36pt

\noindent{References.}

\noindent{1. R. E. Apfel {\it US patent} 4 143 274. (1979)}\\
\noindent{2. R. E. Apfel, S. C. Roy and Y. C. Lo  {\it Phys. 
Rev.} {\bf A31}, 3194 (1985).}\\ 
\noindent{3. H. Ing and H. C. Birnboim  {\it Nucl. Trac. Rad. Meas.} {\bf 8}, 
285 (1984).}\\ 
\noindent{4. R. E. Apfel {\it Nucl. Inst. Meth.} {\bf 179}, 615 (1981).}\\
\noindent{5. R. E. Apfel {\it Nucl. Inst. Meth.} {\bf 162}, 603 (1979).}\\
\noindent{6. F. d'Errico, R. Nath, M. Lamba and K. S. Holland 
{\it Phys. Med. Biol.} {\bf 43}, 1147 (1998).}\\
\noindent{7. B. Roy, B. K. Chatterjee and S. C. Roy {\it Rad. Meas.} 
{\bf 29}, 173 (1997).}\\
\noindent{8. R. E. Apfel and S. C. Roy {\it Rev. Sci. Instrum.} {\bf 54}, 1397 
(1983).}\\ 
\end{document}